\documentclass[reprint,superscriptaddress,notitlepage]{revtex4-1}

\setcitestyle{super, sort&compress}
\usepackage{graphicx}
\usepackage{url}
\usepackage{placeins}
\usepackage{etoolbox}
\usepackage{amsmath}
\usepackage{flafter}
\renewcommand{\vec}[1]{\mathbf{#1}}
 
\begin{document}

\title{Multiple-stable anisotropic magnetoresistance memory in antiferromagnetic MnTe}

\author{D.~Kriegner}
\email[]{dominik.kriegner@gmail.com}
%\thanks{}
\affiliation{Charles University in Prague, Ke Karlovu 3, 121 16 Praha 2, Czech Republic}

\author{K.~V\'yborn\'y}

\author{K.~Olejn\'{\i}k}

\author{H.~Reichlov\'a}

\author{V.~Nov\'ak}

\author{X.~Marti}
\affiliation{Institute of Physics, Academy of Science of the Czech Republic, Cukrovarnick\'a 10, 162 00 Praha 6, Czech Republic}

\author{J.~Gazquez}
\affiliation{Institut de Ci\`encia de Materials de Barcelona ICMAB, Consejo Superior de Investigaciones Cient\'ificas CSIC, Campus UAB 08193 Bellaterra, Catalonia, Spain}

\author{V.~Saidl}

\author{P.~N\v{e}mec}
\affiliation{Charles University in Prague, Ke Karlovu 3, 121 16 Praha 2, Czech Republic}

\author{V.~V.~Volobuev}
\affiliation{Institute of Semiconductor and Solid State Physics, Johannes Kepler University Linz, Altenbergerstr. 69, 4040 Linz, Austria}
\affiliation{National Technical University, Kharkiv Polytechnic Institute, 61002 Kharkiv, Ukraine}

\author{G.~Springholz}
\affiliation{Institute of Semiconductor and Solid State Physics, Johannes Kepler University Linz, Altenbergerstr. 69, 4040 Linz, Austria}

\author{V.~Holy}

\affiliation{Charles University in Prague, Ke Karlovu 3, 121 16 Praha 2, Czech Republic}

\author{T.~Jungwirth}
\affiliation{Institute of Physics, Academy of Science of the Czech Republic, Cukrovarnick\'a 10, 162 00 Praha 6, Czech Republic}
\affiliation{School of Physics and Astronomy, University of Nottingham, Nottingham NG7 2RD, United Kingdom}
\date{\today}

\maketitle

{\bf
A common perception assumes that magnetic memories require ferromagnetic materials with a non-zero net magnetic moment.
However, it has been recently proposed that compensated antiferromagnets with a zero net moment may represent a viable alternative to ferromagnets.\cite{Shick2010,MacDonald2011,Duine2011,Gomonay2014}
So far, experimental research has focused on bistable memories in antiferromagnetic metals.\cite{Park2011,Petti2013,Marti2014,Moriyama2015,Wadley2015} 
In the present work we demonstrate a multiple-stable memory device in epitaxial manganese telluride (MnTe) which is an antiferromagnetic counterpart of common II-VI semiconductors.\cite{Allen1977}
Favorable micromagnetic characteristics of MnTe allow us to demonstrate a smoothly varying antiferromagnetic anisotropic magnetoresistance (AMR) with a harmonic angular dependence on the applied magnetic field, analogous to ferromagnets.
The continuously varying AMR provides means for the electrical read-out of multiple-stable antiferromagnetic memory states which we set by heat-assisted magneto-recording and by changing the angle of the writing field.
We explore the dependence of the magnitude of the zero-field read-out signal on the strength of the writing field and demonstrate the robustness of the antiferromagnetic memory states against strong magnetic field perturbations.
We ascribe the multiple-stability in our antiferromagnetic memory to different distributions of domains with the N\'eel vector aligned along one of the three $c$-plane magnetic easy axes in the hexagonal MnTe film.
The domain redistribution is controlled during the heat-assisted recording by the strength and angle of the writing field and freezes when sufficiently below the N\'eel temperature. 
}

Switching between two memory states in metallic antiferromagnets has been demonstrated in FeRh by taking advantage of its high-temperature ferromagnetic phase,\cite{Marti2014,Moriyama2015}. 
Similar effects were shown in thin-film IrMn tunneling devices utilizing once the exchange-spring with an adjacent ferromagnetic layer\cite{Park2011} or by field-cooling.\cite{Petti2013}
In CuMnAs current induced torques were used to switch the antiferromagnet even without an auxiliary ferromagnetic element and magnetic fields.\cite{Wadley2015}
The two distinct stable states of these antiferromagnetic-metal memories showed distinct resistances which allowed for their electrical read-out.
The physical mechanism of the read-out is ascribed in these devices to the antiferromagnetic AMR.
The interpretation appeals to the general principle formulated by Louis N\'eel in his Nobel lecture that: "Effects in antiferromagnets depending on the square of the spontaneous magnetization should show the same variation as in ferromagnetic substances".\cite{Neel1970}

In ferromagnets, a common characteristic feature of the non-crystalline AMR is its harmonic $\cos2\varphi_B$ and $\sin2\varphi_B$ dependence on the angle $\varphi_B$ between the current and the applied saturating magnetic field when measured with the longitudinal and transverse voltage probes, respectively.
In a conventional Hall-bar geometry, the ratio of these longitudinal and transverse AMR signals scales with the Hall-bar aspect ratio.\cite{McGuire1975}
In antiferromagnets, however, controlling the N\'eel order by external magnetic field is significantly more difficult which has hindered the detection of the characteristic angular dependence of the AMR in antiferromagnets. 

Below we present measurements of the harmonic-function AMR in our Hall-bar devices made in antiferromagnetic semiconductor MnTe and by this directly illustrate the applicability of the above N\'eel's principle on this basic spintronic phenomenon.
Moreover, unlike the previously studied bistable antiferromagnetic-metal memories, our MnTe devices allow us to set by heat-assisted magneto-recording multiple-stable antiferromagnetic memory states.
When removing the writing magnetic fields sufficiently below the N\'eel temperature, the harmonic AMR-like signal is still preserved in the zero-field read-out traces.
As expected for antiferromagnets, these multiple-stable states are not erased by even strong magnetic-field perturbations as we also demonstrate below.  

We use MnTe thin films grown epitaxially on InP(111)A substrates by molecular beam epitaxy.
The quality of our epilayers is evidenced in Fig.~\ref{fig:basic}a which shows a high resolution high angle annular dark field (HAADF) image, showing the individual atomic Te and Mn columns as bright and dimmer spots, respectively.
In contrast to most thin film MnTe studies\cite{Akinaga1993, Janik1995, Hennion2002}, we obtain the hexagonal NiAs bulk structure\cite{Allen1977} ($\alpha$-MnTe) due to the good lattice matching. 
X-ray diffraction measurements shows that this is the only phase present in our films and that $c$-planes are aligned parallel to the substrate surface.
From reciprocal space maps for the 50~nm thick epilayer used in the transport experiments we conclude that the film is relaxed forming a mosaic structure with a block size of $25\pm5$~nm. 
The relative misorientation of the blocks is small with a Gaussian distribution with standard deviation of $0.25\pm0.1^\circ$.
(For more details on the X-ray characterization see Supplementary information (SI) Sec.~I.)  

The semiconducting electronic structure of our $\alpha$-MnTe films is confirmed by optical absorption measurements (see SI Sec.~II) in the visible and mid-infrared range on epilayers deposited on a transparent SrF$_2$(111) substrate.
We performed measurements for layers of different thicknesses, which all show the same onset of the absorption shown in Fig.~\ref{fig:basic}b.
From this we infer an indirect band gap of $1.46\pm0.10$~eV in agreement with previously reported bulk values.\cite{Zanmarchi1967, Allen1977, Ferrer_Roca2000}

Below the N\'eel temperature, which in bulk $\alpha$-MnTe is 310~K\cite{LB_MnTe2000} the magnetic structure consists of ferromagnetically ordered Mn planes which are antiferromagnetically stacked along the $c$-direction.\cite{LB_MnTe2000}
Neutron diffraction experiments\cite{Kunitomi1964,Szuszkiewicz2005} and susceptibility measurements\cite{Komatsubara1963} (see also Fig.~S1b) found that the moments lie within the Mn-planes, as indicated in Fig.~\ref{fig:basic}a.
The moderate N\'eel temperature and the expected weak in-plane magneto-crystalline anisotropy in the hexagonal structure suggest that significant magnetic moment reconfigurations might be achievable at moderate applied magnetic fields.
Low temperature magnetization data measured by the superconducting quantum interference device (SQUID) and shown in Fig.~\ref{fig:basic}c confirm this expectations.
Consistent with the antiferromagnetic order, we observe a zero remanent moment and even at high fields the contribution per Mn atom is only a minor fraction of it magnetic moment of 4.8$\mu_{\rm B}$.\cite{Kunitomi1964,Szuszkiewicz2005}
The moderate exchange energy (N\'eel temperature) and expected weak in-plane anisotropies are reflected in the onset of a sizable magnetic moment at a moderate threshold of $\sim 2$~T in the in-plane magnetic field sweeps.

As shown in SI Sec.~III, the threshold which may be associated with the spin-flop field in the domains moves to lower fields when approaching the N\'eel temperature.
Due to the hexagonal crystal structure and antiferromagnetic moments oriented in the $c$-plane, the system can break into three types of domains in which the N\'eel vector points along one of the three high-symmetry directions.\cite{Komatsubara1963}
In $\alpha$-MnTe single crystals, it was demonstrated that the distribution of these domains can be rearranged by cooling the system through the antiferromagnetic transition in an applied magnetic field.\cite{Komatsubara1963}
All these favorable magnetic characteristics are the basis of our experiments in the MnTe memory devices presented below.

Transport measurements shown in Fig.~\ref{fig:basic}d complete the basic characterization of our thin-film $\alpha$-MnTe samples.
The temperature dependent resistivity shows a peak near the N\'eel temperature associated with critical scattering off spin fluctuations.\cite{Magnin2012}
Note that critical anomalies in our films are consistently observed also in the susceptibility\cite{Komatsubara1963} and lattice parameter measurements (see SI Fig.~S1,~S2).
From the Hall effect measurements we obtained low-temperature hole density of $p=6\times10^{18}$~cm$^{-3}$ due to unintentional doping in our film and corresponding hole mobility of $\mu=43$~cm$^2$/Vs. 

We now proceed to the discussion of the antiferromagnetic AMR and memory functionalities in our MnTe devices.
In Figs.~\ref{fig:amr}a,b we plot the transverse and longitudinal AMRs, defined as {\it AMR}$_\perp(\varphi_B)\equiv\frac{R_{\rm XY}(\varphi_B) - \left<R_{\rm XY}\right>}{\left<R_{\rm XX}\right>} n$ and {\it AMR}$_\parallel(\varphi_B)\equiv\frac{R_{\rm XX}(\varphi_B) - \left<R_{\rm XX}\right>}{\left<R_{\rm XX}\right>}$, were $R_{\rm XY}(\varphi_B)$ and $R_{\rm XX}(\varphi_B)$ are the transverse and longitudinal resistances indicated in the inset of Fig.~\ref{fig:basic}d, $\left<\right>$ denotes averaging over all angles $\varphi_B$, and $n$ is the aspect ratio of our Hall bar.
Measurements are shown after cooling the sample sufficiently below the N\'eel temperature (to 200~K) and then applying a rotating 2~T field.
The curves show a harmonic $\sin2\varphi_B$ ($\cos2\varphi_B$) dependence on the field-angle and the amplitudes of {\it AMR}$_\parallel$ and {\it AMR}$_\perp$ scale with the Hall bar aspect ratio.
Note that the crystalline AMR contribution is negligible in our case (SI Sec. III).
This phenomenology is reminiscent of common AMR traces in ferromagnets in applied saturating magnetic fields.

For comparison, we show in Figs.~\ref{fig:amr}a,b also 2~T AMR curves measured at a low temperature (5~K).
The corresponding traces are anharmonic, show history dependence, have smaller magnitudes than at 200~K, and {\it AMR}$_\parallel$ and {\it AMR}$_\perp$ do not scale with the Hall bar aspect ratio.
The MnTe antiferromagnet becomes stiffer at 5~K and the 2~T field causes only partial reorientation of the spin-axes in the domains around their zero-field direction, reminiscent of ferromagnets in weak fields below the saturation field.

In Fig.~\ref{fig:amr}c we show AMR measurements in which, for each point, we first heated the sample above the N\'eel temperature (to 350~K) and then field cooled (with $B_{\rm FC}$) down to 200~K in a 2~T field of a fixed angle $\varphi_{B,FC}$ and measured the corresponding resistance.
To obtain the data shown in Fig.~\ref{fig:amr}d we continued with the field-cooling down to 5~K then removed the field and took zero-field resistance measurements again at 200~K. (Corresponding data for other read-out temperatures are shown in SI Fig.~S6a.)
Remarkably, we observe similar AMR traces in the two panels only the amplitude of the zero-field AMR is about a factor 2 smaller than in Fig.~\ref{fig:amr}c with the field on.

Note that a 200~K zero-field AMR of a comparable amplitude to the one seen in Fig.~\ref{fig:amr}d is also obtained when field-cooling from 350~K down to only 200~K.
Crossing the N\'eel temperature in the heat-assisted magneto-recording helps the efficiency of the writing process.
It is not necessary, however.
Setting a stable AMR trace of the form of Fig.~\ref{fig:amr}d, only with a factor 2 smaller amplitude, is also possible by field-cooling from 200~K down to low temperatures and then performing the zero-field read-out measurement back at 200~K (see SI Fig.~S7).
This is consistent with the ability, seen in Figs.~\ref{fig:amr}a,b, to control the antiferromagnetic state by the 2~T field even if below, but not too far from the N\'eel temperature. 

We point out that Fig.~\ref{fig:amr} not only demonstrates a continuous harmonic-function AMR in an antiferromagnet but it also shows a multiple-stability of states in our MnTe memory device.
Next we explore how the amplitude of the memory read-out signal depends on the strength of the writing field and test the limits for erasing the multiple-stable states by turning the magnetic field back on.
In Fig.~\ref{fig:fc}a we plot the AMR amplitude, defined as $\frac{R_{\rm XY}(45^\circ) - R_{\rm XY}(-45^\circ)}{\left<R_{\rm XX}\right>}  n$, obtained when setting the states by field-cooling at $\varphi_{B,FC}\pm45^\circ$ from 350~K down to 5~K and measuring the resistance during the temperature down-sweep with the field on.
Fig.~\ref{fig:fc}b shows zero-field AMR amplitudes measured during the subsequent up-sweep in temperature. 

For the range of writing field amplitudes from 0.5 to 2~T we were able to set the multiple-stable memory states with the zero-field AMR of the harmonic form seen in Fig.~\ref{fig:amr}d.
The signal disappears for all traces at $\sim285$~K, which is consistent with the N\'eel temperature of thin $\alpha$-MnTe films\cite{Przedziecka2005} and confirms the antiferromagnetic origin of the read out signal.
The amplitude of the AMR signal scales with the amplitude of the writing field, apart from the region between 0.5 and 1~T where the AMR signal changes sign, and does not saturate at the maximum applied writing field of 2~T.
Before discussing the origin of these features we complete in Figs.~\ref{fig:fc}c,d the description of experimental data, namely of the measurements testing the robustness of the multiple-stable memory states under magnetic field perturbations.

In Fig.~\ref{fig:fc}c we replot the zero-field AMR amplitude obtained from the temperature up-sweeps after setting the states in 2~T writing fields.
We compare the trace with analogous measurements which only differ in an additional magnetic field exposure of the memory at 5~K after the writing and before starting the temperature up-sweep read-out measurement.
The additional exposure comprises application of rotating in-plane and out-of-plane magnetic fields of a 1 or 2~T amplitude.
We see that neither of these two amplitudes of the disturbing field applied at any angle is sufficient to fully erase the memory.
After the exposure, the multipe-stable states maintain their characteristic harmonic AMR form of the read-out signal (see SI Fig.~S6) with only the amplitude being partially reduced, as shown in Fig.~\ref{fig:fc}c. 

We performed similar attempts to erase our antiferromagnetic memory at 200~K.
The results, shown in Fig.~\ref{fig:fc}d, illustrate that at this elevated temperature the 2~T magnetic field is sufficient to fully erase the memory states.
This is consistent with measurements in Figs.~\ref{fig:amr}a,b in the rotating 2~T field at 200~K, where the observed AMR traces had a character of AMR in ferromagnets under saturating magnetic fields. 
A 1~T field, on the other hand, is not sufficient to erase the memory at 200~K; it only reduces the amplitude of the read-out signal, as seen in Fig.~\ref{fig:fc}d.

In Fig.~\ref{fig:fc}e we illustrate in more detail the stability of our antiferromagnetic memory in fields, which are insufficient to erase it.
We explore how the states set by cooling in 2~T writing fields of angles $\varphi_{B,FC}=\pm45^\circ$, corresponding to the extrema in the {\it AMR}$_\perp$ read-out signal, are disturbed at 5~K by a 1~T field rotating in the sample plane. 
By taking the resistance measurements with the field on we observe a partial reorientation of the antiferromagnetic spin-axis, reflected in the varying resistance signal. 
The variations are, however, smaller than the difference between the zero-field resistances of the two extrema. 
Moreover, if at each given angle of the disturbing magnetic field we remove the field and repeat the resistance measurement, we see that the original zero-field state almost fully recovers.

An additional proof of only partial reorientations of moments in the antiferromagnetic domains by the applied rotating field at 5~K is given in Fig.~S8 in the SI. We observe that the ratio of the amplitudes of {\it AMR}$_\parallel$ and {\it AMR}$_\perp$ depends on the previous field-cooling protocol and generally does not scale with the Hall bar aspect ratio. This is in contrast to measurements at 200~K in the 2~T rotating field (Fig.~\ref{fig:amr}a,b) where the AMR shows the scaling with the Hall bar aspect ratio, independent of the history.  

We now proceed to the discussion of the origin of the observed multiple-stable aniferromagnetic memory states and the corresponding zero-field harmonic-function AMR.
When cooling the system in an applied magnetic field, the domain distribution freezes at temperatures sufficiently below the N\'eel temperature, as illustrated in Fig.~\ref{fig:model}a. 
The freezing occurs when the thermal fluctuations no longer allow the different domains to flip their spin-axis from one to the other metastable direction.

The writing magnetic field applied at a certain angle during the cooling makes the easy axis closest to the field normal more favorable (see sketches in Fig.~\ref{fig:model}b). 
Due to the finite temperature, however, the other two easy axis directions can be also populated with the statistics given by the Boltzmann distribution and by the relative strength of the terms contributing to the total energy of the system for the given domain size, namely the exchange energy, the anisotropy energy, and the Zeeman energy. 

When the Zeeman term dominates, i.\,e. for very large magnetic fields or domain sizes, one of the three easy directions would be strongly favored within a 60$^\circ$ interval of the writing field angles, and the other directions for the other respective 60$^\circ$ intervals. 
The read-out AMR signal in this case would take the form of a step-like trace with three distinct memory states.
Upon reducing the Zeeman contribution, the AMR trace undergoes a transition into a continuous function whose amplitude gradually decreases and whose shape approaches the harmonic $\sin2\varphi_{B,FC}$ function of the writing field angle.
Since our measured signals plotted in Fig.~\ref{fig:amr}d show a weak but clearly detectable deviation from the harmonic $\sin2\varphi_{B,FC}$ form we can infer the typical domain size from fitting the model to the measured data.

The distribution of the three different easy-axis domains and the corresponding AMR signals are calculated using the antiferromagnetic Stoner-Wohlfarth model\cite{Bogdanov1998}. 
The results of the modelling, plotted in Fig.~\ref{fig:amr}d, reproduce well the experimental behavior.
In the model, the exchange energy was estimated from the N\'eel temperature and the magnetic moment found in neutron diffraction experiments.\cite{Kunitomi1964,Szuszkiewicz2005}
The anisotropy energy was then inferred assuming a spin-flop field of $\sim 2$~T.
(For more details on the model see SI Sec.~V.)

From the fitting we obtain that all three domains are at least partially occupied with the highest inequality of 2:1:1 and 2:2:1 for the cases when the writing field is perpendicular or parallel to one of the three easy axes, respectively.
The moments easy orientation is found to be along $\left<10\bar10\right>$ directions as indicated in Fig.~\ref{fig:basic}a.
The domain size we find in the fitting is approximately 20~nm, which is in a good agreement with the mosaic block size obtained from the X-ray diffraction results. 
From this we conclude that the magnetic domain size is limited by the structural block size and does not change significantly with the writing conditions. 

Under the assumption of a constant domain size we use the model to also estimate the dependence of the amplitude of the zero field read-out signal on the strength of the writing field. 
As shown in Fig.~\ref{fig:model}d, the model predicts an initial quadratic increase of the read-out AMR signal followed by a tendency towards saturation at higher fields.
It broadly agrees with the measured data apart from the negative AMR signal seen in the experiment at low writing fields. 
We attribute this negative signal to domain walls which may have some uncompensated moments. 
Unlike the compensated moments in the antiferromagnetic domains which tend to align perpendicular to the applied field, these uncompensated moments prefer a parallel alignment. 
Since the uncompensated moments only lead to a linear gain in the Zeeman energy the effect is diminished at larger fields by the canting of the moments in the antiferromagnetic domains which yields an energy gain quadratic in the applied magnetic field.

To summarize, we have reported spintronic memory functionalities in an antiferromagnetic counterpart of common II-VI compound semiconductors.
Favorable magnetic characteristics of our $\alpha$-MnTe epilayers allowed us, on one hand, to evidence the direct analogy between AMR in antiferromagnets and ferromagnets, both in its basic phenomenology and in the utility as an electrical detection tool of the ordered magnetic moments.
On the other hand, our work also highlights the unique potential of antiferromagnets for spintronics.
In our MnTe devices we demonstrated that the harmonic-function AMR persists even after removing the setting magnetic field and that the multiple-stable memory states cannot be erased by strong magnetic field perturbations when sufficiently below the antiferromagnetic transition temperature.
These magnetic memory characteristics are unprecedented in ferromagnets. 

\section*{Methods}

{\bf Sample preparation.} 
$\alpha$-MnTe was grown by molecular beam epitaxy on (111) oriented InP and SrF$_2$ substrates using elemental Mn and Te sources and 
substrates temperatures in the range of 370 to 450$^\circ$C. 
2D growth was achieved in both cases as judged from the streaked RHEED patterns observed during growth.
The orientation of our layers on the substrate is ($0001$)[$10\bar10$]$_{\rm MnTe}$$\parallel$ ($111$)[$11\bar2$]$_{\rm sub}$.
For structural characterization we carried out cross-sectional scanning transmission electron microscopy (STEM) in the high angle annular dark field (HAADF) imaging mode and x-ray diffraction studies with CuK$\alpha_1$ radiation. 
STEM-HAADF images were acquired with a NION UltraSTEM, equipped with a NION aberration corrector and operated at 100 kV. 
In the images the intensity scales approximately with the atomic number (Z) squared.\cite{Nellist1999}
Cross sectional STEM specimens were prepared using a FEI Nova 200 Dual-Beam SEM/FIB focused ion beam.

\begin{figure}
\centering
\includegraphics[width=0.7\textwidth]{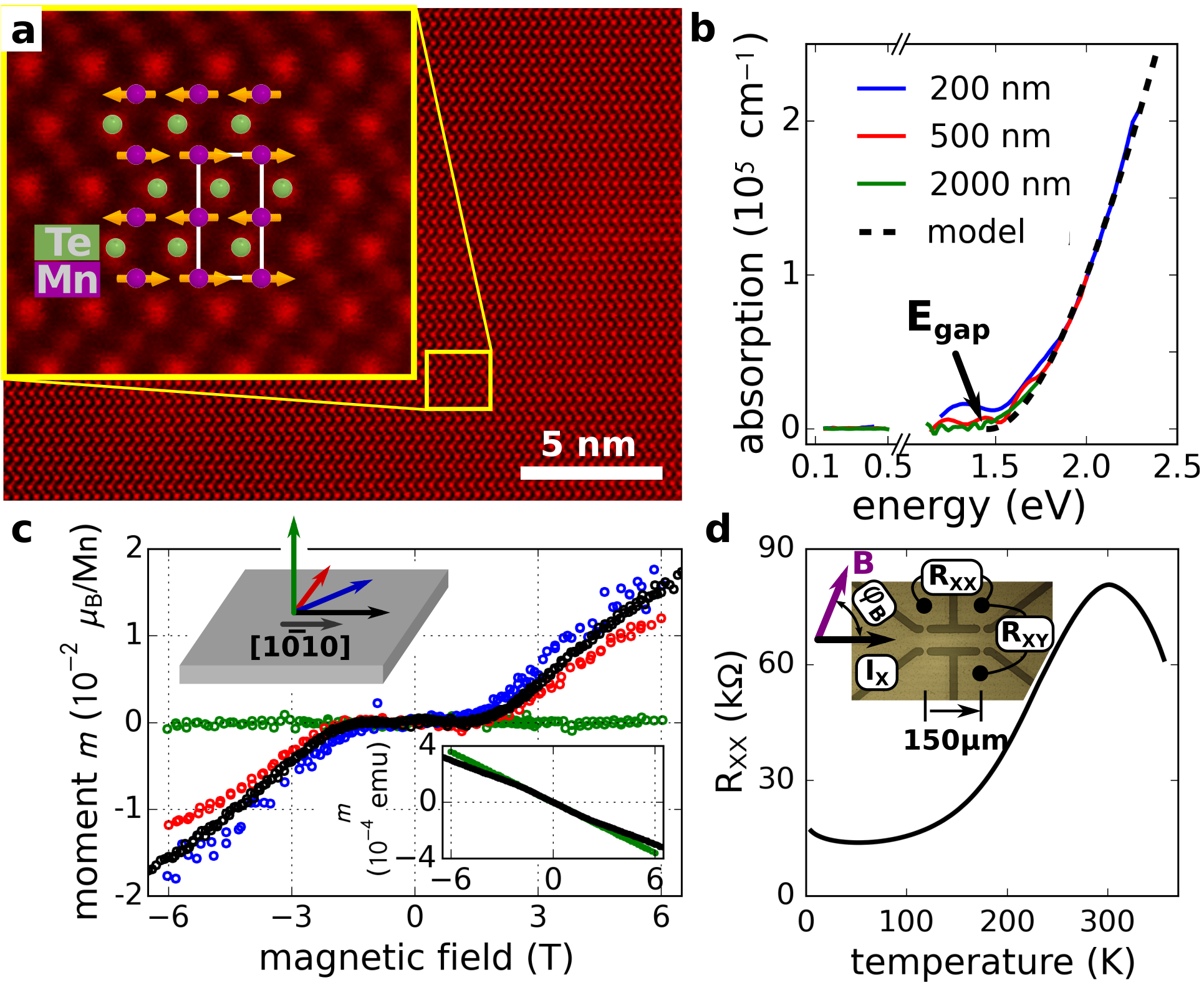}
\caption{\label{fig:basic} {\bf MnTe thin films properties.} 
a) Cross sectional high resolution high angle annular dark field (HAADF) image taken along the $\left[1\bar210\right]$ zone axis resolving the individual atomic columns of Mn (magenta) and Te (green). 
The inset shows the atomic positions including the Mn magnetic moments in the ordered antiferromagnetic state and the unit cell size is indicated by a white line. 
b) Absorption coefficient of MnTe in the mid-infrared and visible spectral range extracted from transmission measurements for three different film thicknesses.  
c) SQUID magnetic field sweeps for different field directions at 5~K. 
The slope of the out-of-plane measurement was subtracted from all the measured traces.
An inset shows the as measured moments dominated by the diamagnetism of the substrate. 
d) Temperature dependent resistance measurement showing a peak around the transition temperature. 
An inset shows the Hall bar geometry and the definition of the magnetic field angle $\varphi_B$.}
\end{figure}

\begin{figure}
\centering
\includegraphics[width=0.7\textwidth]{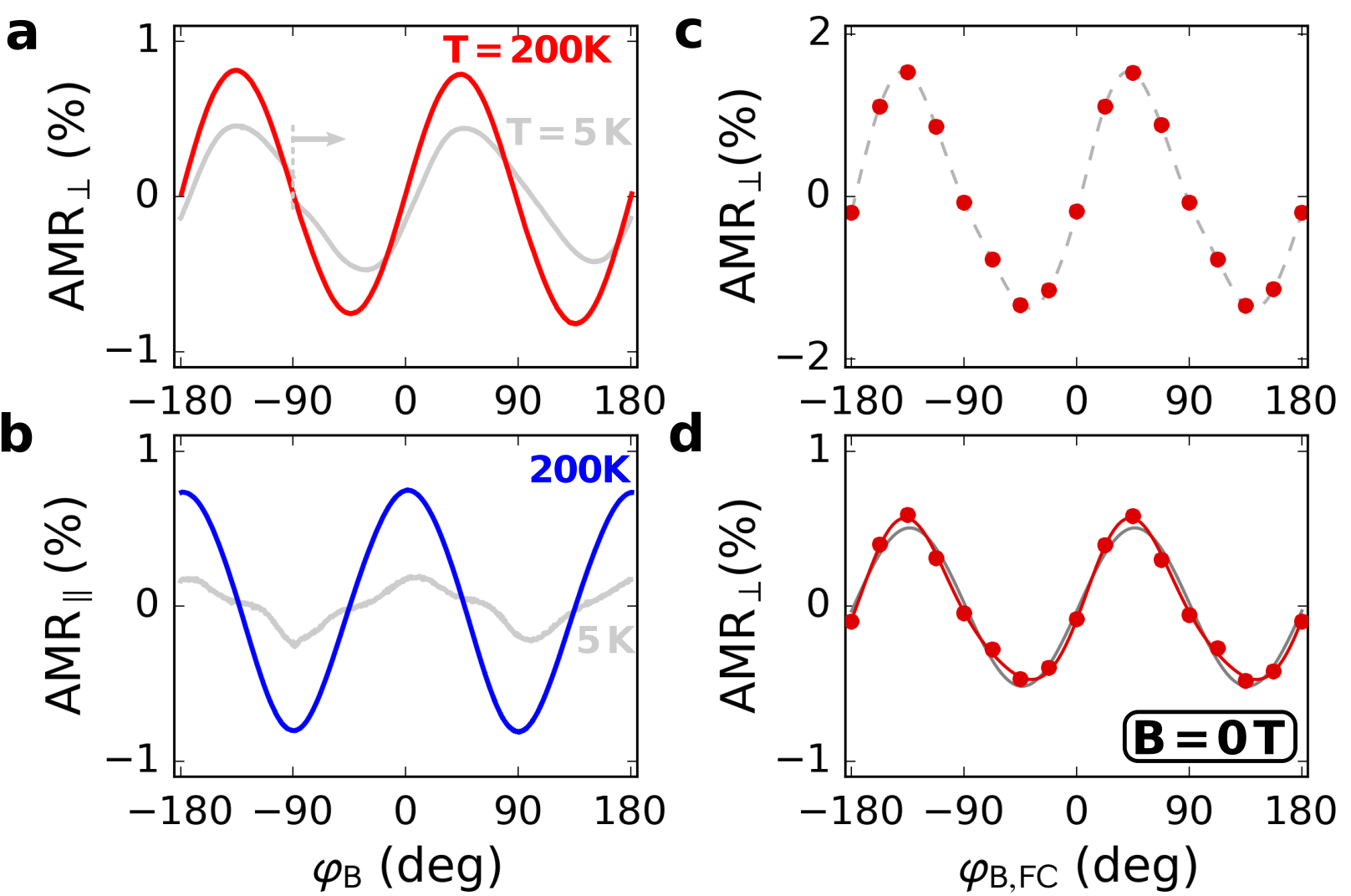}
\caption{\label{fig:amr} {\bf Antiferromagnetic anisotropic magnetoresistance and multiple-stable memory.} 
a,b) Transverse (red) and longitudinal (blue) AMR measurements at 200~K and rotating in-plane 2~T field. Analogous measurements at 5~K are shown in grey. 
Arrow in panel a) indicates the initial angle and the direction of  rotation. c) Transverse AMR measured at 200~K after cooling from 350~K in a 2~T magnetic field applied at an angle $\varphi_{B,FC}$ and with the field kept on. 
d) Zero-field transverse AMR obtained after continuing with the field-cooling down to 5~K then removing the field and taking zero-field resistance measurements at 200~K. The red line shows the fit of the model calculations described in the text, while the grey line is a $\sim\sin 2\varphi_{B,FC}$ fit.}
\end{figure}

\begin{figure}
\centering
\includegraphics[width=0.85\textwidth]{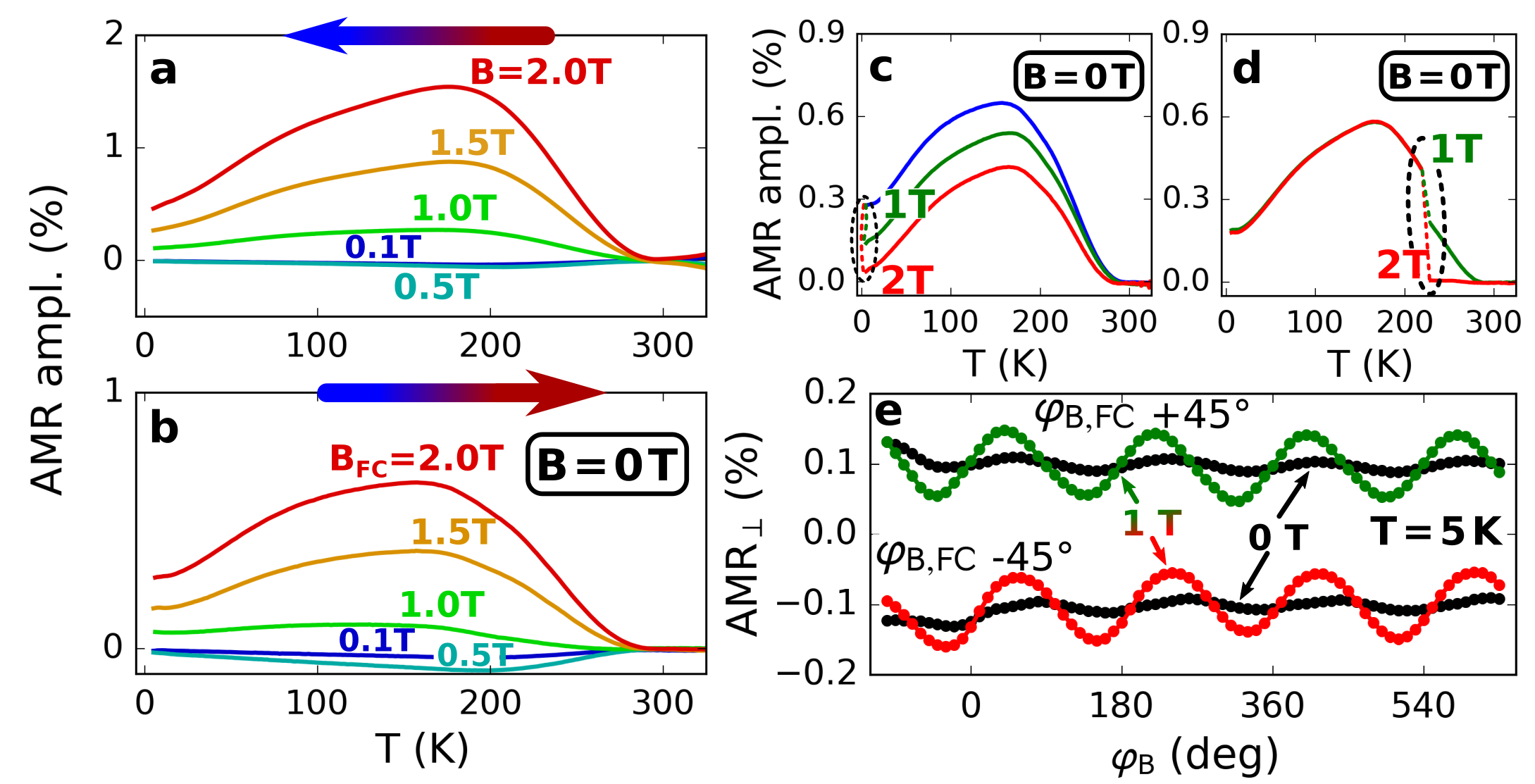}
\caption{\label{fig:fc} {\bf Dependence on writing field and robustness against field perturbations.} 
a) Temperature and writing field-strength dependence of the AMR amplitude for the temperature down-sweep with the writing field kept on.
b) Temperature and writing field-strength dependence of the zero-field AMR amplitude for the temperature up-sweep.
c) Zero-field AMR amplitudes for the temperature up-sweep after exposing the memory at 5~K to 1~T (green) and 2~T (red) perturbing fields rotated both in-plane and out-of-plane. 
Blue curve represents the unperturbed reference measurement obtained with $B_{\rm FC}=2$~T. 
d) Same as c) for the field perturbations applied at 200~K. 
e) The stability of two memory states set by the heat-assisted magneto-recording from 350~K with a 2~T writing field applied at angles $\varphi_{B,FC}=\pm45^\circ$ tested at 5~K by a rotating 1~T field. 
At every angle $\varphi_B$ of the perturbing 1~T field, red/green curves correspond to the read-out resistance measurements with the field on while black lines are obtain after removing the perturbing field at each $\varphi_B$.}
\end{figure}

\begin{figure}
\centering
\includegraphics[width=0.7\textwidth]{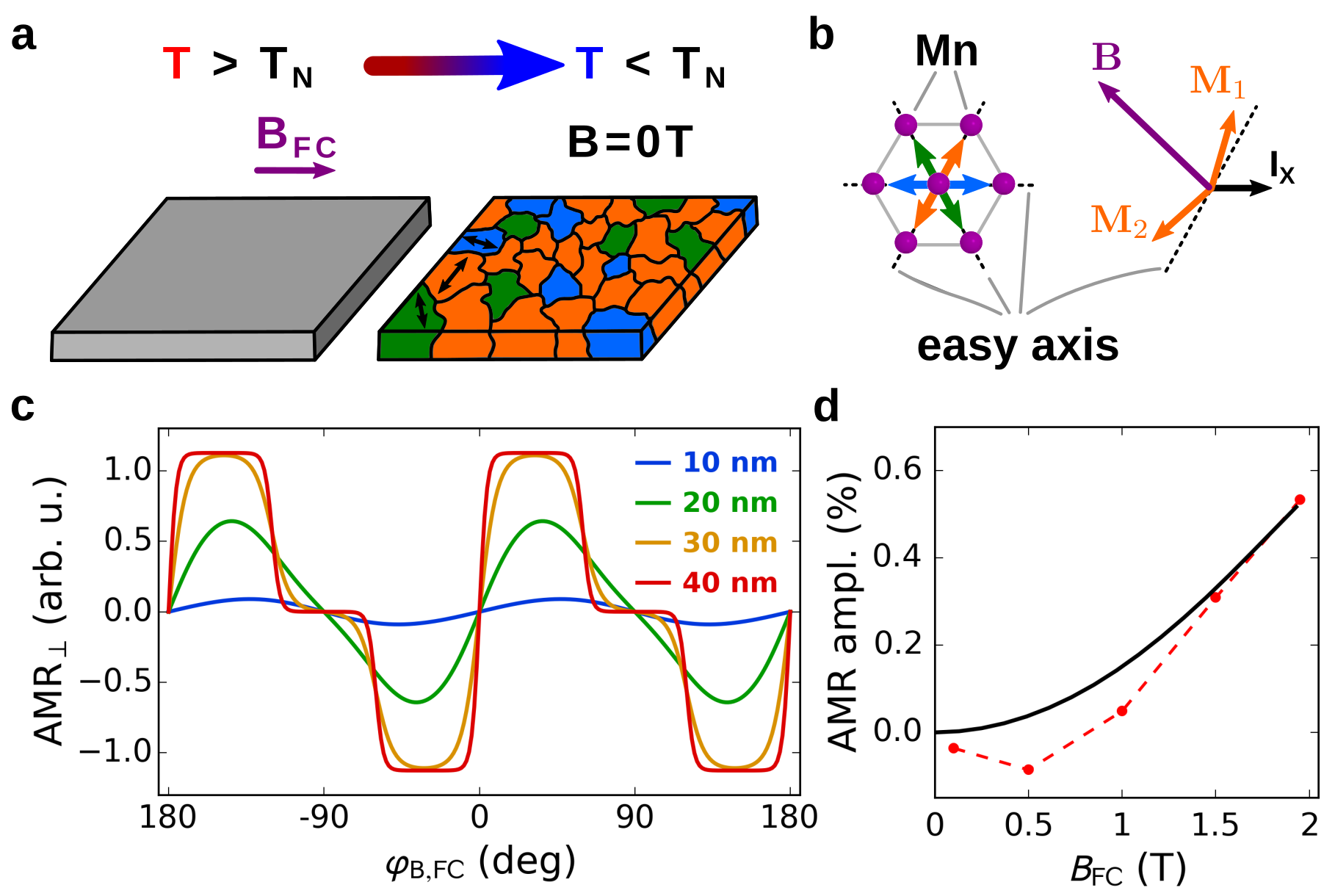}
\caption{\label{fig:model} {\bf Modeling of the multiple-stable states.}
a) Schematics of setting the multiple-stable states, each with a distinct domain distribution, by the heat-assisted magneto-recording across the N\'eel temperature. 
b) Schematics of the two antiferromagnetic spin sublattices which tend to align perpendicular to the applied magnetic field and to cant towards the field; schematics of the three in-plane easy axes in the hexagonal $\alpha$-MnTe. c) Modelling of the distribution of the three domains in the multiple-stable memory states and of the corresponding transverse AMR signal for different size of the domains. 
d) Comparison of the model prediction (black line) and experimental dependence at 200~K (red points) of the read-out AMR amplitude on the strength of the writing magnetic field.}
\end{figure}

{\bf Magnetometry} measurements were performed on 2000~nm thick MnTe films with area of $\sim11$~mm$^2$ and 500~$\mu$m thick InP substrate in a Quantum Design SQUID magnetometer using reciprocating sample option (RSO) for increased measurement sensitivity.
Magnetic field sweeps were recorded for various crystallographic directions while sweeping from negative to positive and vice versa showing no hysterisis.
From all magnetic field sweeps a negative slope obtained from the fit to the hard-axis out-of-plane sweep seen in the inset in Fig.~\ref{fig:basic}c was subtracted to correct for signal coming primarily from the diamagnetic substrate.
Temperature dependent susceptibility measurements were taken in magnetic field of 0.5~T.

{\bf Optics}
The absorption coefficient and thus band gap properties were determined by fitting the transmission spectra (measured at 300~K) with an analytical expression\cite{Swanepoel1983} for the transmission of a thin absorbing film on a transparent substrate.

{\bf Processing and transport measurements}
For transport measurements Hall bars were defined in 50~nm thick films grown on InP along the [$10\bar10$] direction by electron-beam lithography and wet etching using a H$_3$P0$_4$/H$_2$O$_2$/H$_2$0 mixture as etchant.
Contacts were made by soldering with pure In and 4-point measurement technique was employed to avoid possible contribution of contact resistances.
Complementary Corbino-disk devices were fabricated by deposition of Au rings.
The transport measurements were performed in an Oxford Instruments vector field cryostat capable of 2~T in arbitrary directions and up to 6~T in the current direction.
A constant DC current density of 2000~A/cm$^2$ was applied during all measurements.

\begin{acknowledgments}
We thank A.B.~Shick and M.~Veis for fruitful discussions and M.~Mary\v{s}ko for technical support.
This work is supported from the Austrian Science Fund (FWF): J3523-N27 and SFB-025 IRON, from the European Research Council (ERC) Advanced Grant No. 268066, from the Ministry of Education of the Czech Republic Grant No.~LM2011026, and from the Grant Agency of the Czech Republic Grant No.~14-37427G.
J.\,G. acknowledges the Ramon y Cajal program (RYC-2012-11709).
Scanning transmission electron microscopy research (J.G.) was supported through a user project at Oak Ridge National Laboratory’s Center for Nanophase Materials Sciences (CNMS), Scientific User Facilities Division, BES-DOE. 
\end{acknowledgments}

%\bibliographystyle{naturemag}
%\bibliography{refs}

\pagebreak 
\onecolumngrid

\clearpage
\renewcommand{\thetable}{S\arabic{table}}%
\renewcommand{\thefigure}{S\arabic{figure}}%
\setcounter{figure}{0}
\makeatletter
\apptocmd{\thebibliography}{\global\c@NAT@ctr 26\relax}{}{}
\makeatother
{\huge \centering{Supplementary Informations\\ --\\ Multiple-stable anisotropic magnetoresistance memory in antiferromagnetic MnTe}}

\section{MnTe thin films - growth and basic properties}

\begin{figure}
\centering
\includegraphics[width=0.6\textwidth]{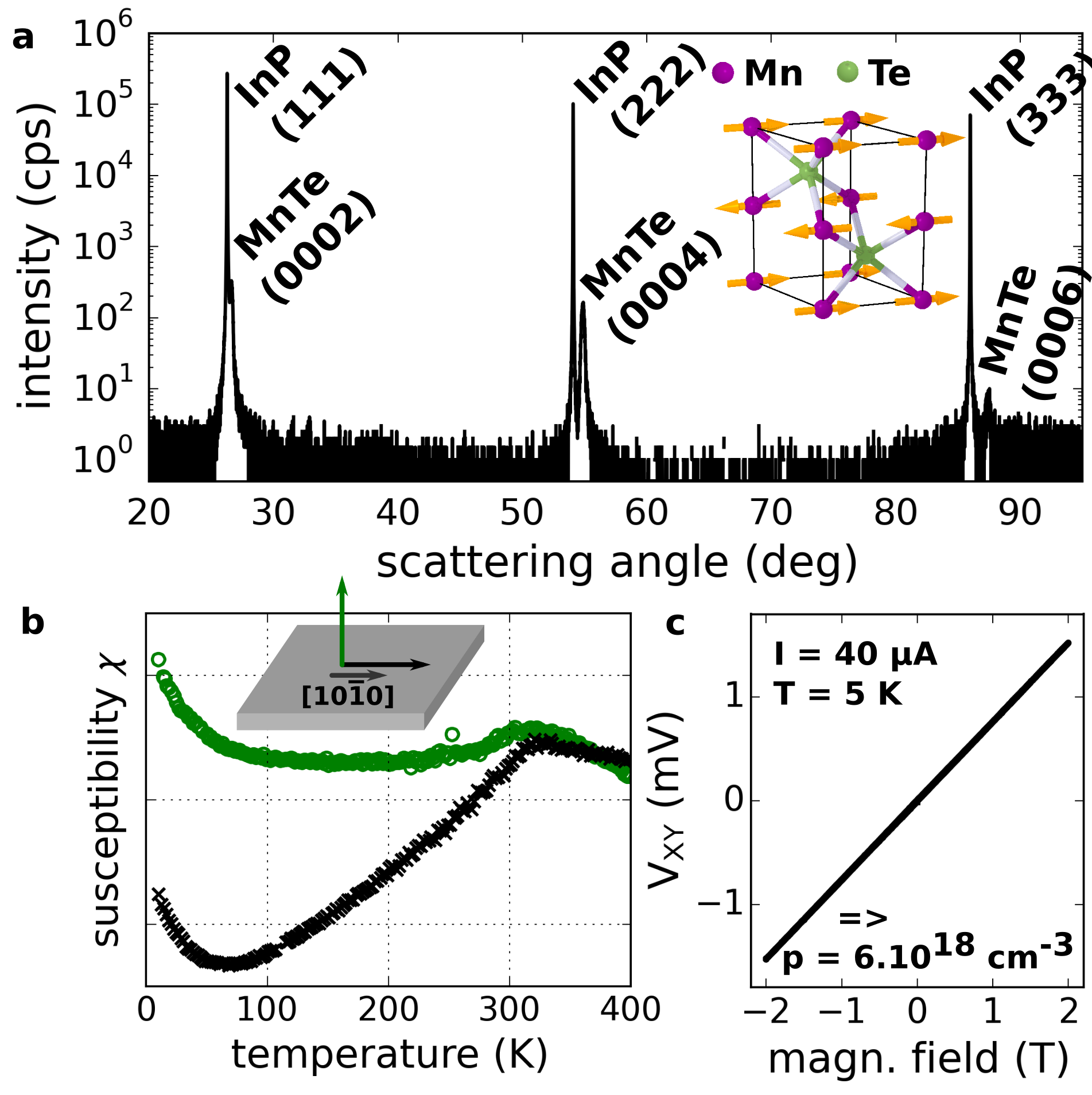}
\caption{\label{fig:basicS} {\bf MnTe thin film properties.} 
a) Xray diffraction scan of a 50~nm thick MnTe film grown on InP(111) showing only phases of the hexagonal bulk $\alpha$-MnTe phase which is sketched in the inset. 
b) Magnetic susceptibility of $\alpha$-MnTe thin films measured for in-plane and out of plane field orientation. 
The drop of the susceptibility in the in-plane configuration is a result of the in-plane moment orientation. 
c) Hall measurement in order to determine the carrier density and mobility.}
\end{figure}

\begin{figure}
\centering
\includegraphics[width=0.65\textwidth]{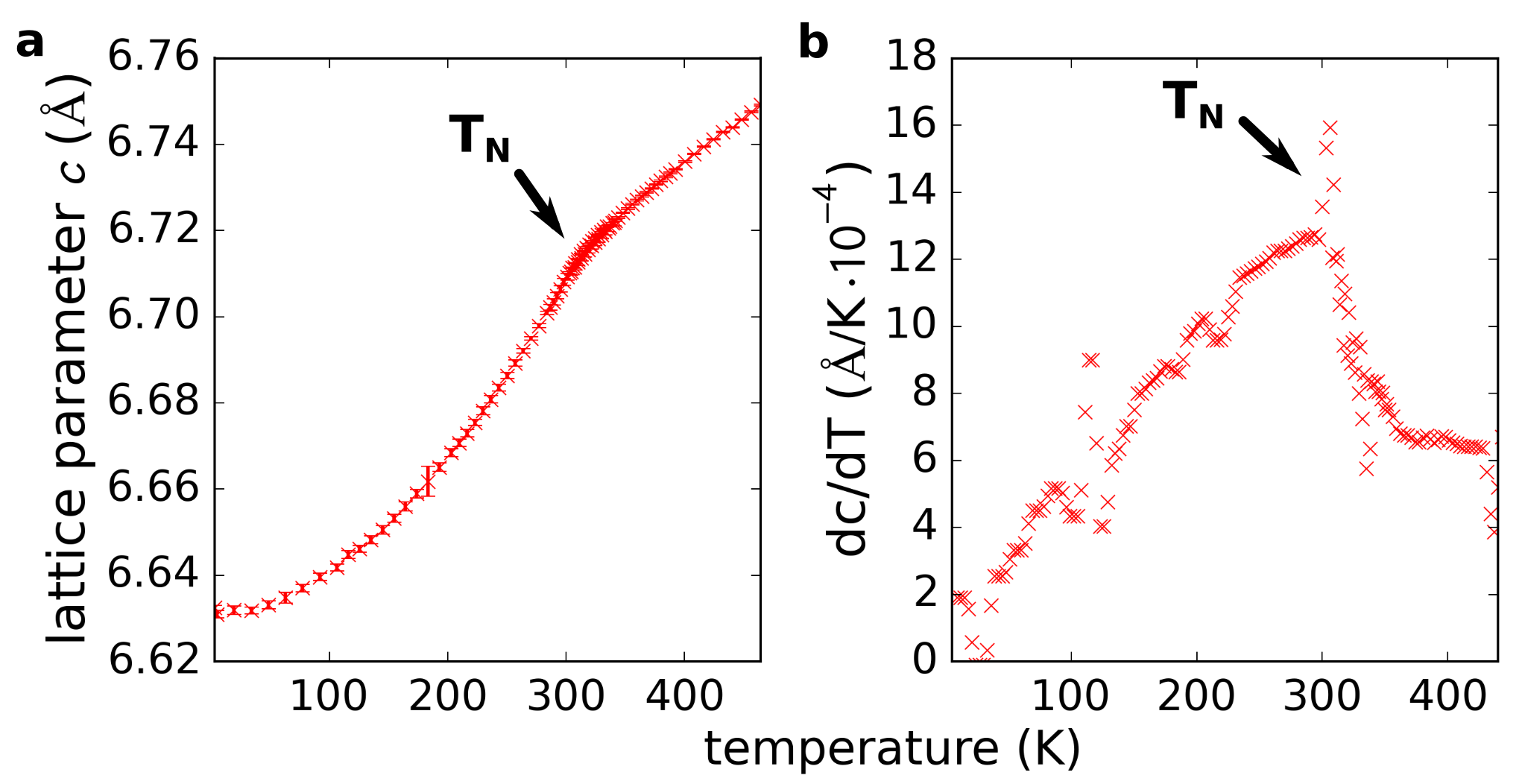}
\caption{\label{fig:xrdtemp} {\bf temperature dependent $c$ lattice parameter.}
a) The out of plane lattice parameter shows a anomaly in its thermal expansion properties at the AFM transition temperature.
b) In the derivative of the lattice parameter the anomaly is even more pronounced.
Measurements were performed on a 2000~nm thick $\alpha$-MnTe grown on InP.
The N\'eel temperature $T_N$ is indicated by arrows.}
\end{figure}

MnTe thin films were grown on InP(111) and SrF$_2$(111) substrates for transport and optical investigations, respectively.
Growth was performed by molecular beam epitaxy using elemental sources at a substrate temperature of 370 to 450$^\circ$C.
Using X-ray diffraction we find that the thin films grow in the hexagonal NiAs bulk phase ($\alpha$-MnTe) and that no other phases are present (Fig.~\ref{fig:basicS}a).
In magnetometry measurements, although the signal is dominated by the temperature independent diamagnetism of the substrate, we are able to detect the transition temperature around $\sim300$~K (Fig.~\ref{fig:basicS}b) from a peak in the susceptibility. 
Furthermore we find that the magnetic moments are oriented in the sample plane as evident from the drop in susceptibility which is more pronounced for an in-plane magnetic field, equivalent to the report in Ref.~21 (main text).
Temperature dependent diffraction experiments reveal an anomaly of the out of plane lattice parameter at the transition temperature\cite{Greenwald1953} (Fig.~\ref{fig:xrdtemp}), confirming the transition temperature found by the magnetometry measurements.

The semiconducting properties of our $\alpha$-MnTe thin films are clearly seen in the Hall measurement shown in Fig.~\ref{fig:basicS}c, which shows a linear dependence of the Hall voltage on an out of plane applied magnetic field. 
We find a hole carrier density of $p = 6 \times 10^{18}$~cm$^{-3}$ with a mobility of $\mu = 43$~cm$^2/$Vs at $T = 5$~K.

\subsection{Epitaxial orientation determined by X-ray diffraction}

\begin{figure}
\centering
\includegraphics[width=0.7\textwidth]{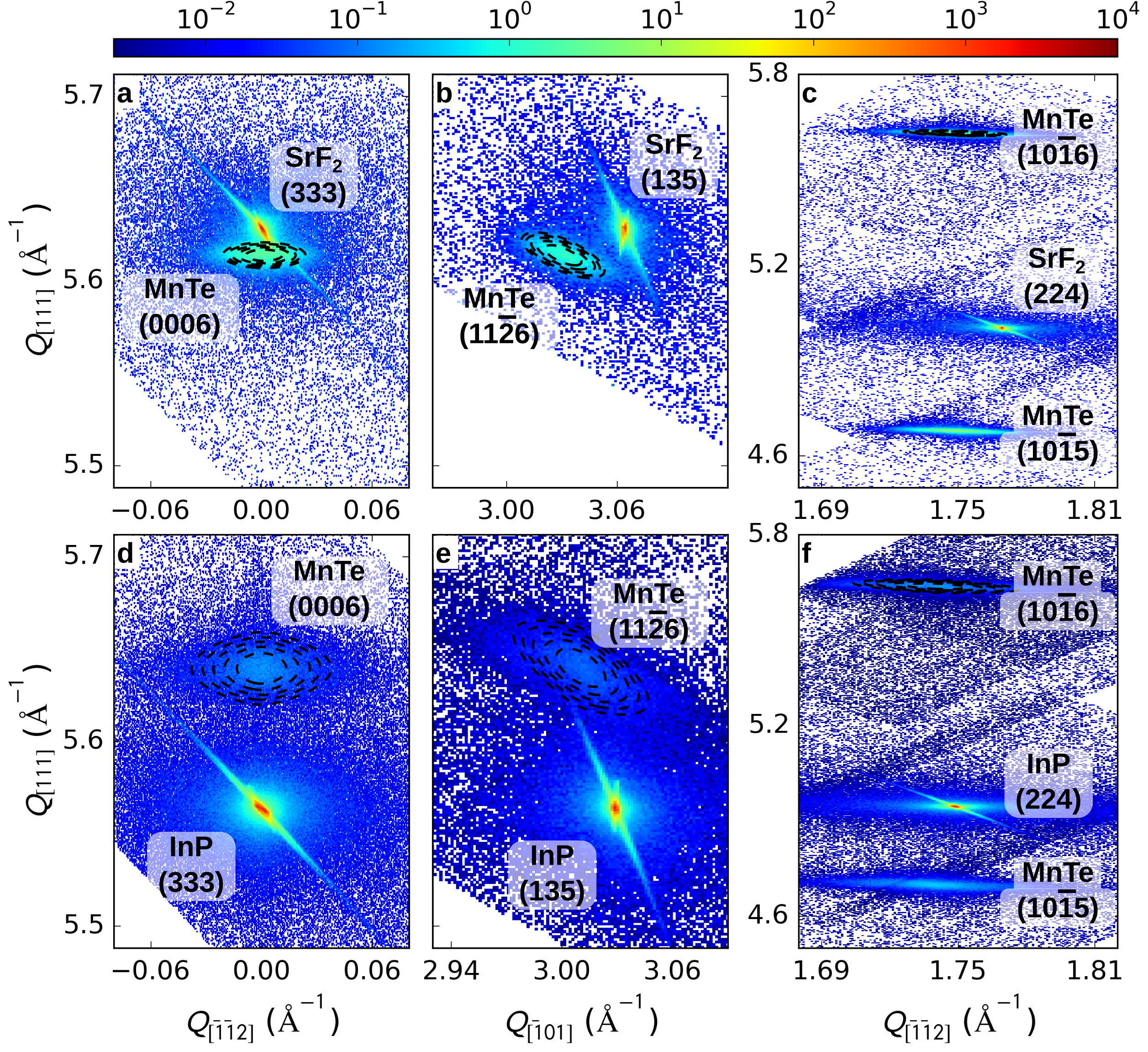}
\caption{\label{fig:rsms} {\bf X-ray diffraction reciprocal space map measurements.}
The color plot shows the experimental data with labeled peaks from the SrF$_2$/InP substrate and $\alpha$-MnTe film.
The shown measurements are for 200~nm MnTe on SrF$_2$ in panels (a-c) and 50~nm on InP in (d-f).
Dashed contour lines mark the fit with the mosaic block model.}
\end{figure}

Even though the mismatch between $\alpha$-MnTe and the InP(111) or SrF$_2$(111) substrates is below 1\% the films grow relaxed for the investigated thickness range of 50 to 2000~nm. 
From the peak positions in the reciprocal space maps we find lattice parameters of $a=4.1708$~\AA\ $c=6.6860$~\AA\ for 50~nm MnTe grown on InP.
Further we are able to determine the in-plane epitaxial orientation of the hexagonal $\alpha$-MnTe lattice on the cubic substrate.
For both types of substrates the orientation of the $c$-planes (0001)\footnote{We use the Miller-Bravais indices ($hkil$) with $i=-h-k$ to denote hexagonal directions/reciprocal space points.} is parallel to the (111) planes of the cubic substrate and the in-plane [$10\bar10$] direction of $\alpha$-MnTe corresponds to the [$11\bar2$] direction of the substrate.
Figure~\ref{fig:rsms} shows reciprocal space maps measured for $\alpha$-MnTe thin films grown on both types of substrate.
With a mosaic block model\cite{Pietsch2004} (dashed contour lines in Fig.~\ref{fig:rsms}) we are able to describe the peak shape of the diffraction signal for both types of samples (grown on InP(111) or SrF$_2$(111)).
For the sample used in the transport measurements (grown on InP) the lateral and vertical blocksize is found equal around $25\pm5$~nm and a Gaussian rotation distribution with standard deviation of $0.25\pm0.1$~degree was used for the simulations.
The sample used in optics (grown on SrF$_2$) can be described by blocks of same lateral size, however, with $50\pm2$~nm vertical block size and Gaussian rotation distribution with standard deviation of $0.15\pm0.05$~degree.

\FloatBarrier
\section{Optical characterization}

\begin{figure}
\centering
\includegraphics[width=0.7\textwidth]{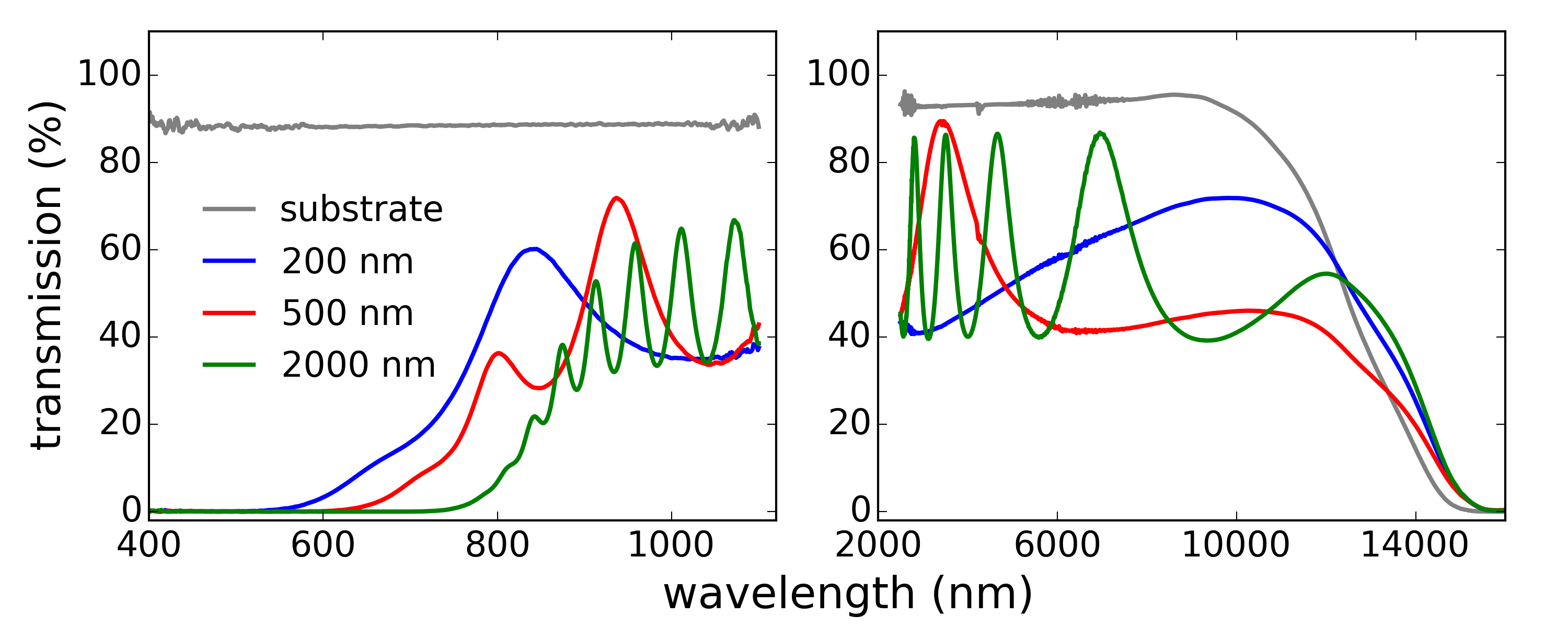}
\caption{\label{fig:optics} {\bf Optical characterization of MnTe by transmission measurements.} 
Transmission spectra of the $\alpha$-MnTe thin films grown on SrF$_2$(111) in the visible (left) and mid-infrared (right) spectral regions. 
Data for samples with various thicknesses allow us to extract a value of absorption coefficient shown in the main text.}
\end{figure}

We determine the spectral dependency of the absorption coefficient for samples with different film thickness and find low absorption in a wide spectral range as typical for a semiconductor.
Below a 700--800~nm all the films show an onset of strong absorption due to excitation of carriers across the band gap. 
The used samples were grown on SrF$_2$, which is transparent in the full visible spectral range and up to 10~$\mu$m in the infrared (Fig.~\ref{fig:optics}ab) ruling out any influence of the substrate.
Due to the film thickness we observe Fabry-P\'erot oscillations in the transmission data at wavelength above the onset of the strong absorption. 
By modeling (see Ref.~26 in main text) these oscillations considering the sample thickness we extract the thickness independent absorption coefficient from the measurement. 
The spectral dependence in Fig.~1b (main text) is typical for an indirect bandgap material. 
The size of the band gap at room-temperature is found as $E_g = 1.46 \pm 0.10$~eV. 

\FloatBarrier
\section{Magnetic field threshold at elevated temperatures}

\begin{figure}
\centering
\includegraphics[width=0.7\textwidth]{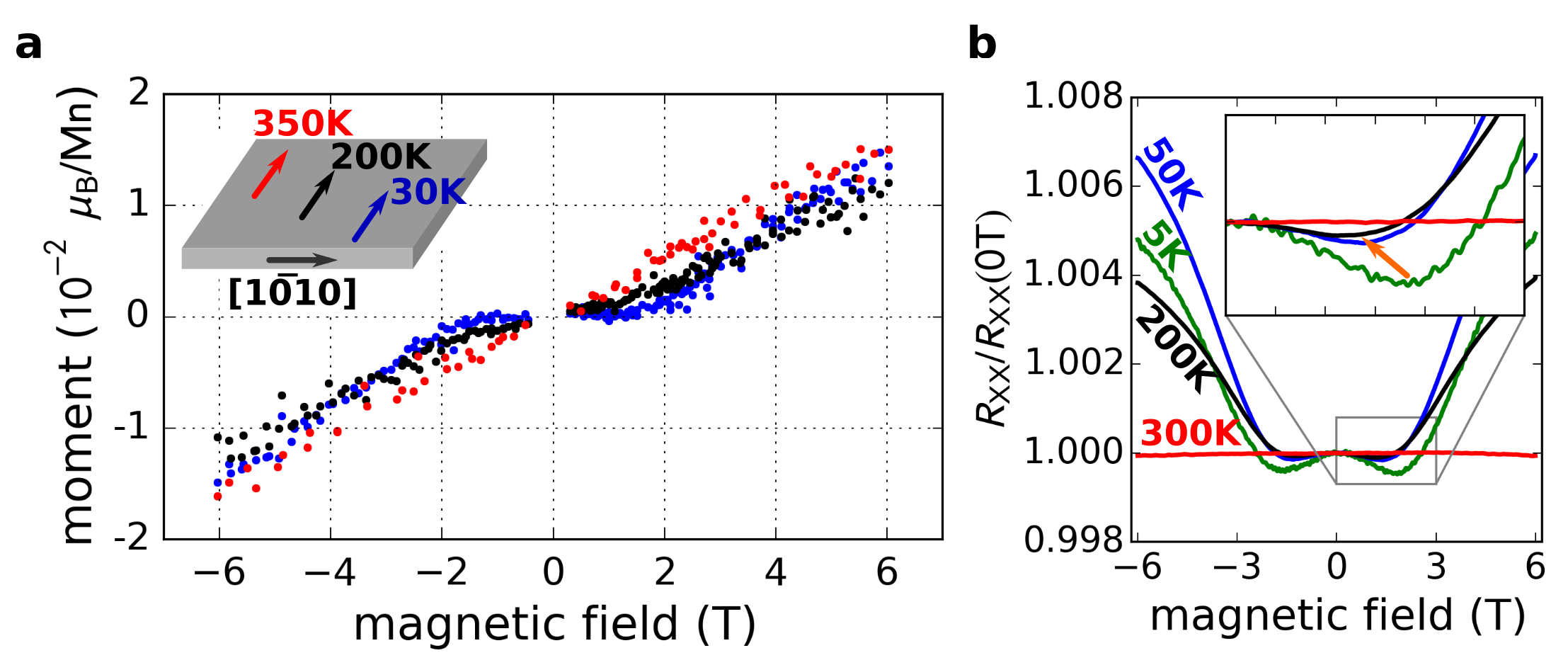}
\caption{\label{fig:squidkink} {\bf Magnetometry measurements and correlation with resistance at elevated temperatures.} 
a) Magnetic moment in in-plane magnetic field sweeps recorded at different temperatures. 
The temperature independent diamagnetic contribution of the substrate was subtracted as explained in the main text. 
The kink around 2~T gradually disappears at higher temperatures.
b) Transport measurements in Corbino geometry at various temperatures show that an analogous kink can be observed in the resistance which via the crystalline AMR effect is sensitive to the orientation of the magnetic moments.}
\end{figure}

From the magnetometry data shown in the main text we conclude that the low temperature onset of a sizable magnetic moment occurs at a $\sim 2$~T in-plane field which may be associated with the spin-flop field in the antiferromagnetic domains.
For higher temperatures the magnetometry measurements are shown in Fig.~\ref{fig:squidkink}a.
These measurements were also correlated with in-plane field sweeps in a Corbino geometry.
The longitudinal magnetoresistance shows a similar change of slope at low temperatures.
This feature remains clearly visible at higher temperatures and allows us to conclude that the critical field needed for a reorientation of the moments decreases at higher temperatures. 
At 200~K the kink is sufficiently below $2$~T to allow manipulation of the N\'eel vector by application of a 2~T field.
From the Corbino geometry we further can estimate the size of the crystalline anisotropic magneto-resistance (AMR), which was found to be below 0.03\% at 200~K and 2~T.
It is therefore a negligible contribution in the transport data presented in the main text.

\FloatBarrier
\section{Stability of the frozen states}

\begin{figure}
\centering
\includegraphics[width=0.65\textwidth]{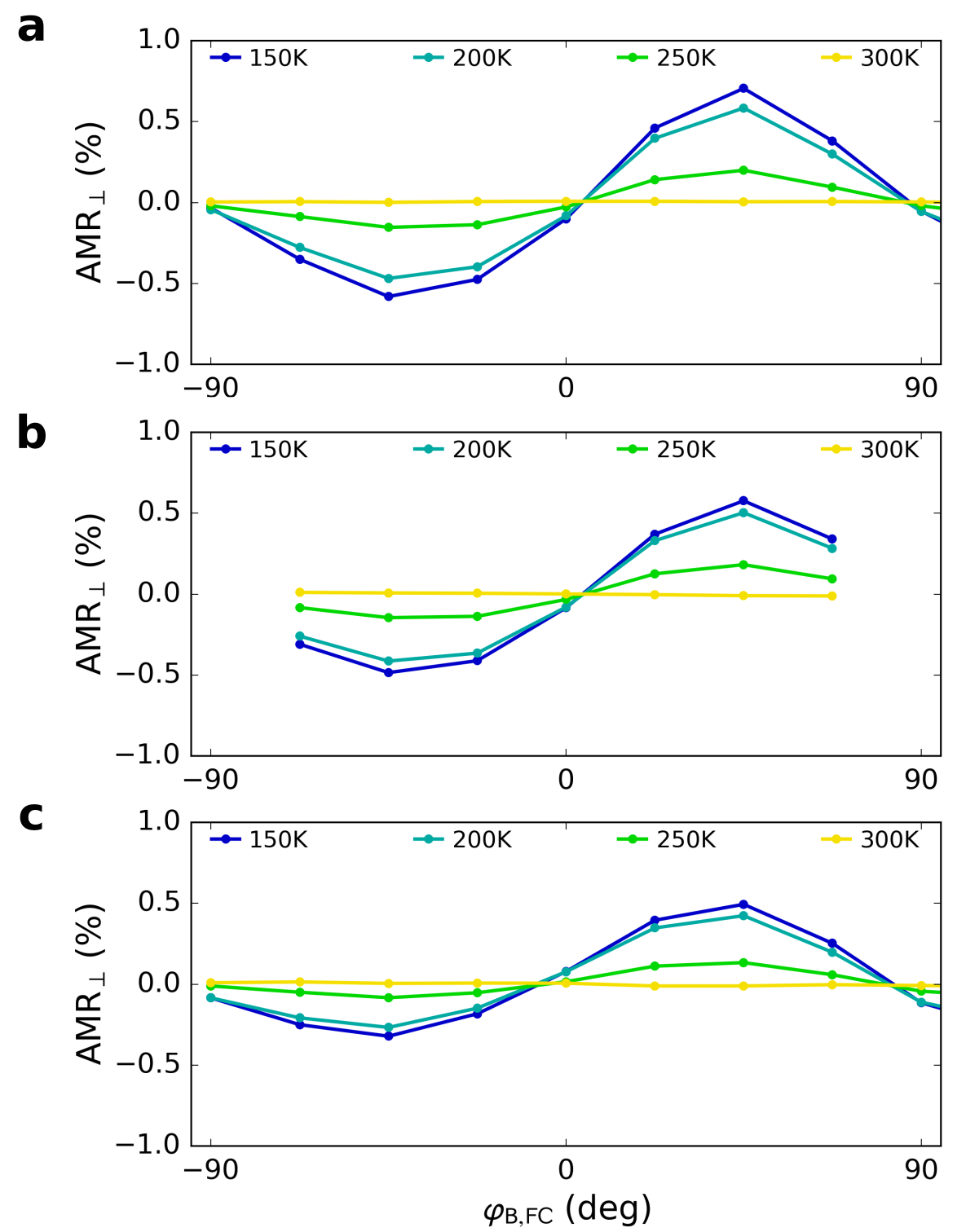}
\caption{\label{fig:sumstability} {\bf Stability of the AMR signal against field rotations.} 
a) The zero field AMR signal induced by field cooling in various directions is shown for different temperatures.
Below 300~K there is clear semi harmonic signal due to the frozen moment orientation.
The same signal with equal symmetry is found in a measurement after the magnetic field of a) 1~T and b) 2~T magnitude was rotated in three perpendicular planes at 5~K.
Although the amplitude of the AMR is slightly reduced these field rotations can not destroy the frozen state.}
\end{figure}

In the main text we show the zero field AMR signal at 200~K (Fig.~2d) as it results after field cooling from above N\'eel temperature to 5~K.
Figure~\ref{fig:sumstability}a shows the equivalent data for different temperatures. 
The AMR signal is maximum at 150~K and reduces to zero at the magnetic order transition.
To test the stability of this signal with respect to magnetic field at low temperature we performed field rotations at 5~K.
Although the zero field AMR signal is decreased as can be seen in Fig.~\ref{fig:sumstability}b,c it can not be fully erased by neither 1~T nor 2~T rotations.
This is in contrast to equivalent rotations at 200~K, where a rotation of a 2~T field (above the spin-flop field) is able to destroy the frozen state as seen in main text Fig.~3d.

The fact that with 2~T at 200~K we are in a different regime as at 5~K is also seen in the ratio of the {\it AMR}$_\parallel$ to {\it AMR}$_\perp$ amplitudes. 
Only at 200~K and 2~T the respective amplitudes scale with the Hall bar aspect ratio.
In the other cases (5~K: 1 and 2~T, and 200~K 1~T) the data show a dependence on the field cooling direction and strongly deviate from the ideal ratio of 1. 

\begin{figure}
\centering
\includegraphics[width=0.45\textwidth]{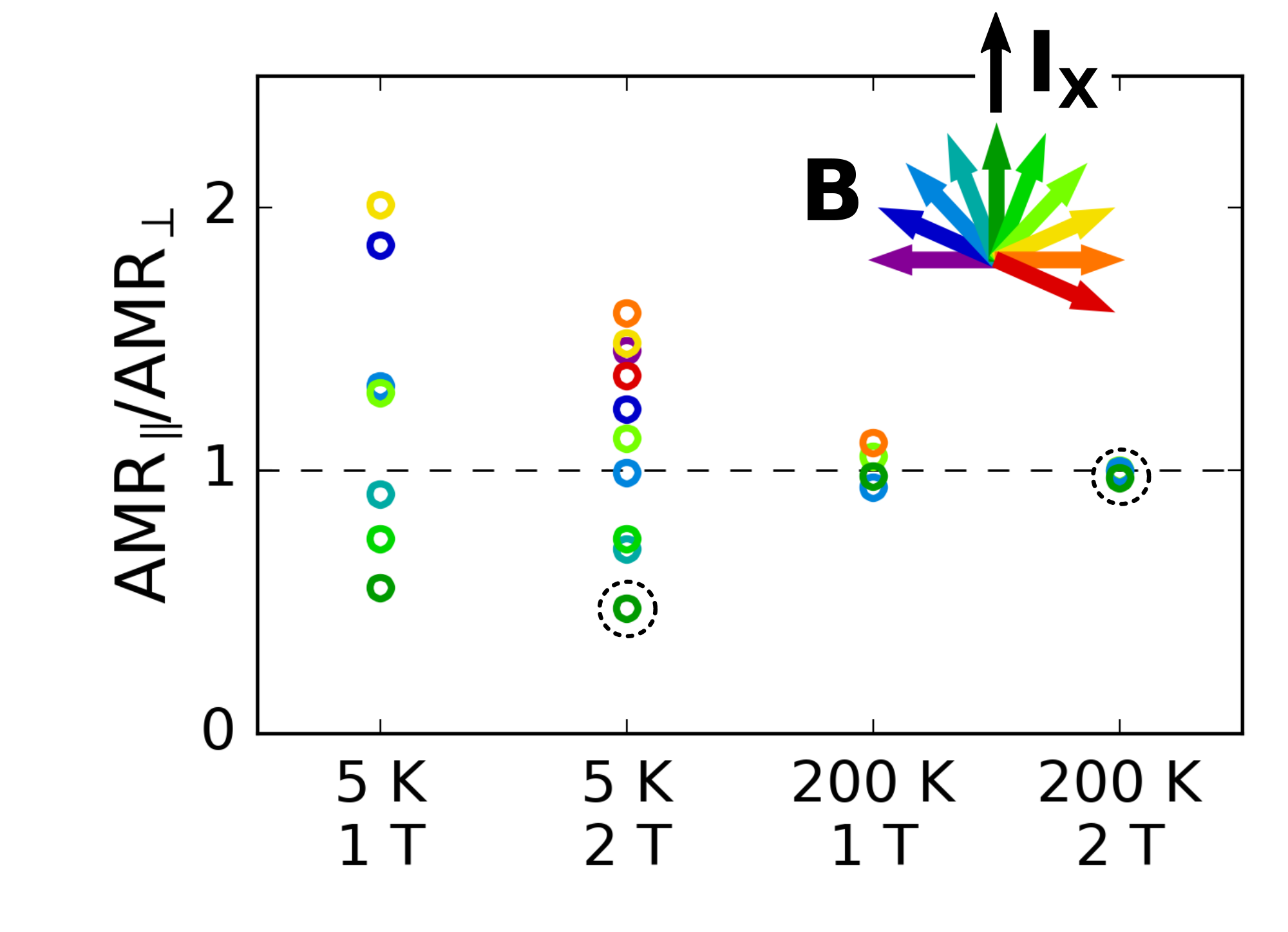}
\caption{\label{fig:amrratio} {\bf {\it\bf AMR}$_\parallel$ to {\it\bf AMR}$_\perp$ ratio for different field cooling directions.} 
The ratio of the AMR signal extracted from the longitudinal and perpendicular contacts shows strong dependence on the field cooling direction at 5~K. 
At 200~K and 2~T the ratio is independent of the history of the sample indicating that 2~T are above the critical field for reorientation of the moments.
Dashed circles mark the data points extracted from the curves shown in Fig.~2a,b (main text).}
\end{figure}

\begin{figure}
\centering
\includegraphics[width=0.65\textwidth]{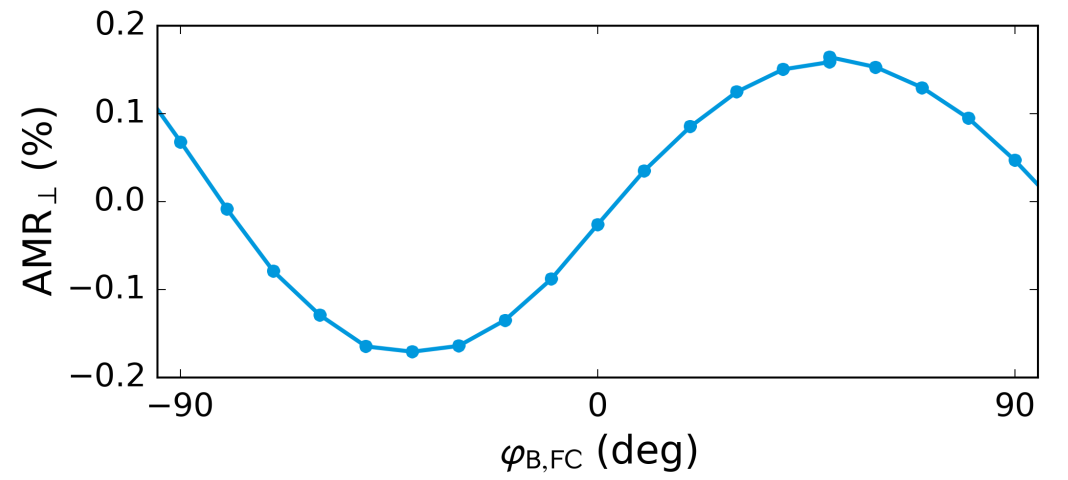}
\caption{\label{fig:fc200_5} {\bf Magneto-recording by cooling from 200 to 5~K.} 
{\it AMR}$_\perp$ data recorded in zero field at 200~K after field cooling from 200 to 5~K in a 2~T applied field.}
\end{figure}

Due to the fact that at 200~K we are able to overcome the spin-flop field with a 2~T in-plane field we are also able to perform magneto-recording when cooling in field from 200 to 5~K. 
Corresponding data are shown in Fig.~\ref{fig:fc200_5} and clearly demonstrate that at 200~K AMR traces are found in zero field although the N\'eel temperature was never overcome. 

\FloatBarrier
\section{Multiple-stability and anistropic magnetoresistence model}

\begin{figure}
\centering
\includegraphics[width=0.3\textwidth]{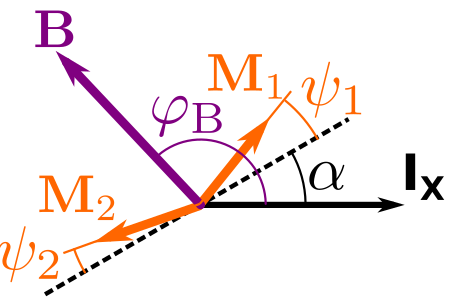}
\caption{\label{fig:sw} {\bf Sketch of the parameters in the Stoner--Wohlfarth model.} 
The angles of the magnetic moments of the two sub-lattices deviate slightly from the easy axis direction upon the application of an magnetic field.
For the modeling relevant vectors and angles are indicated.}
\end{figure}

In order to describe the domain population after the field cooling and the resulting AMR signal we use the Stoner--Wohlfarth model. 
We apply it to each magnetic sub-lattice of $\alpha$-MnTe in order to describe the anisotropy of the material and we introduce a term accounting for the exchange energy.
We restrict ourself to a 2D description which is sufficient to describe the field cooling in different in-plane magnetic fields.
Every magnetic domain consists of two equal ferromagnetic sub-lattices which have opposing magnetization vectors $\vec M_{1,2}$. 
When a magnetic field $\vec B$ is applied in the plane of the sample the magnetic moments might cant with respect to each other and thereby tilt away from the easy axis direction.
The relevant angles are defined with respect to the current or easy axis direction as sketched in Fig.~\ref{fig:sw}.

The total energy $E$ per sample volume $V$ of a single domain is given by the sum of exchange, Zeeman, and magneto--crystalline anisotropy energies:
\begin{equation}
E/V = J_{\rm ex} \vec{\hat M}_1 \cdot \vec{\hat M}_2 - \vec B \cdot (\vec M_1 + \vec M_2) + E_{\rm MAE}(\vec{\hat M}_1) + E_{\rm MAE}({\vec M}_2)
\end{equation}
where $J_{\rm ex}$ is the exchange constant, $E_{\rm MAE}$ is the magneto-crystalline anisotropy funtion, and a hat (''\ $\hat{}$\ '') denotes a unit vector.
Using the angles denoted in Fig.~\ref{fig:sw} and the exchange field defined as $B_{\rm ex} = J_{\rm ex}/M$ with $M = \left|\vec M_{1,2}\right|$ this can be rewritten as
\begin{align}
E/V =& - M B_{\rm ex} \cos( \psi_1 + \psi_2 ) - M B \left[ \cos\left(\varphi_{\rm B} - (\alpha + \psi_1)\right) - cos\left(\varphi_{\rm B} - (\alpha - \psi_2)\right) \right] \nonumber \\
     & + E_{\rm MAE}(\psi_1) + E_{\rm MAE}(\psi_2)
\end{align}
The exchange field we obtain from the N\'eel temperature as $B_{\rm ex} = k_{\rm B}/\mu_{\rm B} T_{\rm N}$, where we use the Boltzmann constant $k_{\rm B}$ and the Bohr magnetron $\mu_{\rm B}$.
The magneto-crystalline anisotropy energy density we express as
\begin{equation}
    E_{\rm MAE}(\psi) = K_{\rm MAE} \sin^2(3\psi),
\end{equation}
in order to describe the six-fold in-plane symmetry of the hexagonal material.
$K_{\rm MAE}$ has units of an energy density and can be obtained from the spin flop field $B_{\rm SF}$ and the exchange field by $K_{\rm MAE} = 12 M B_{\rm SF}^2/B_{\rm ex}$.

From the descibed model we can obtain the energy gain $\Delta E_i$ for the three different easy axis directions $\alpha_i$ when a magnetic field is applied at a certain angle. 
For this we find the arrangements of the moments by minimizing the energy for every field direction.
Using the obtained energy gain for a certain sample volume $V$ in comparison with the thermal energy at the moment of the freezing in the Boltzmann distribution we obtain the relative population $w_i$ of the different domains.
\begin{equation}
    w_i = \frac{e^{\Delta E_i/(k_{\rm B} T_{\rm F})}} {\sum_{j=1}^3 e^{\Delta E_j/(k_{\rm B} T_{\rm F})}}
\end{equation}

This population numbers we use to calculate the transversal and longitudinal AMR signal which is proportional to $\sum_{i=1}^3 w_i \sin(2\alpha_i)$ and $\sum_{i=1}^3 w_i \cos(2\alpha_i)$, respectively. 
Calculating the expected zero field AMR for different domain sizes one can see how the field cooling efficiency is increasing for bigger domains leading to stronger AMR (Fig.~4c in the main text).
When the domains are getting big enough so that almost only one domain is populated the AMR signal is getting step-like.
The results of our calculations in comparison with the experimental data, which show a transitional behavior between the harmonic AMR and the step-like signal, are shown in the main text Fig.~2d.
By adjusting the domain size and easy axis direction as well as amplitude of the AMR a nearly perfect agreement with the experimental shape could be found.
The easy axis direction is found to be along the $\left<10\bar10\right>$ directions and the domain size was found around 20~nm similar to the domain size determined by the mosaic block model.
The variation of the AMR amplitude with field cooling strength also qualitatively agrees with the experimental observations.

%\bibliographystyle{naturemag}
%\bibliography{refs}

\end{document}